# Observations of time delayed all-optical routing in a slow light regime

B. S. Ham

The Center for Photon Information Processing and the Graduate School of Information and Telecommunications,
Inha University
253 Younghyun-dong, Nam-gu, Incheon 402-751, S. Korea
bham@inha.ac.kr; Phone: +82-32-860-8423; Fax: +82-32-865-0480

**Abstract:**
We report an observation of a delayed all-optical routing/switching phenomenon based on ultraslow group velocity of light via nondegenerate four-wave mixing processes in a defected solid medium. Unlike previous demonstrations of enhanced four-wave mixing processes using the slow light effects, the present observation demonstrates a direct retrieval of the resonant Raman-pulse excited spin coherence into photon coherence through coherence conversion processes.
PACS numbers: 42.50.Gy, 42.65.Ky,

Quantum coherent control of light by using another light has been studied intensively for the speed control of the light for the last several years in atomic vapors [1,2], cold atoms [3], defected solid crystals [4,5], semiconductors [6], optical fibers [7], photonic crystals [8,9], and silicon ring resonators [10]. The physics of the light speed control comes from the Kramers-Kroing relations, in which dispersion and absorption profile is coupled each other in a form of an integral. Especially slowing down the group velocity of a traveling light pulse has been the heart of research in optical and quantum memories. The slow light phenomenon, therefore, can act as a key role to quantum optical information processing.

Ultrahigh-speed all-optical switching technologies have been studied for THz optical data processing [11,12]. Optical switching devices should come along with an optical buffer memory, but all-optical buffer memory techniques [13] are primitive and far behind in comparison with other optical switching techniques. Even though information bandwidth of an optical fiber reaches up to 100 THz, optical switching devices in fiber-optic communications network still rely on electrical-optical and/or optical-electrical conversion processes, which are limited to a few GHz. Because the processing speed of an electronic CPU is at most a few GHz, the overall performance of an optical switching device must be limited by the slow counterpart of electronics. Therefore, a combined research on both all-optical switch and all-optical buffer memory gives a great benefit to potential applications.

For the last two decades electromagnetically induced transparency (EIT) [14] has been studied in various types of optical media and showed enhanced nonlinear optical processing using a slow-light phenomenon [15]. It is well known that EIT results in non-absorption resonance owing to destructive quantum interference even in an optically thick medium when two optical fields interact with each other. Needless to say the main advantage of EIT is in the group velocity control with easy-hand modifications of absorption spectrum keeping absorption free at resonance frequency. This means that the ultra-slow group velocity of light can be obtained without severe loss of energy.

It should be noted that the transparency window of EIT is strongly correlated with atomic spin coherence. This light-matter correlation is understood as dark state polariton [16]. As an ultimate case of the slow light, stop light or a zero group velocity has been studied for an optical quantum memory [4,17-19]. In this case, the reversal quantum mapping process between optical photons and atomic coherence must be understood carefully. Unlike an atomic medium, rare-earth doped solids utilize inhomogeneously broadened spins as well as optical population shelving [20].

In a population shelved limit [4,12,20], where the optical decay rate is much smaller than applied optical Rabi frequency, the atomic population transfer by spontaneous decay processes between two low lying spin states via an excited optical state can be negligible. Moreover, inhomogeneously broadened spins render photon-echo [21] like phenomenon, so multiple-bit quantum optical memory is possible [22]. In a non-slow light regime for a traveling light pulse, the EIT-induced spin-optical coherence conversion process has been demonstrated via nondegenerate four-wave mixing processes [23]. Unlike slow-light based wave mixing dynamics in atomic media [24-26], where the wave mixing process is enhanced owing to existing slow light due to EIT interwound wave mixing, a direct coherence conversion from the slow-light enhanced spin coherence to optically coherent photons via nondegenerate four-wave mixing processes is demonstrated. In addition a delayed all-optical routing phenomenon is also discussed in the EIT-based slow light regime.

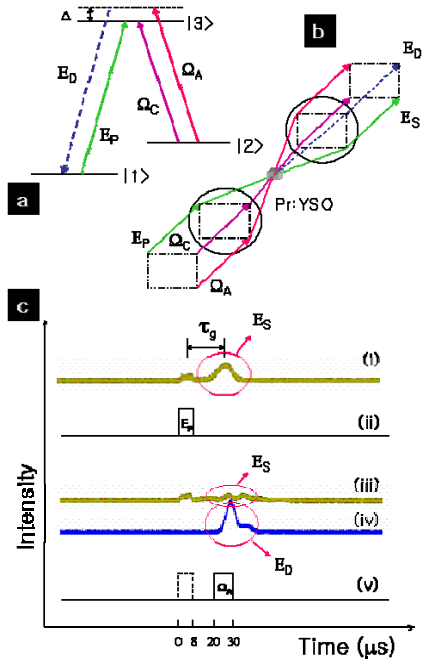

**Figure 1 Experimental observation of delayed all-optical routing. a**, Energy level diagram. **b**, Schematic diagram of laser pulse propagation. **c.** (i) EIT-induced slow light, (ii) Pulse sequence for (i), (iii) and (iv) delayed all-optical routing, and (v) pulse sequence for (iii) and (iv).

In Fig. 1a the ground states |1> and |2> are chosen from the hyperfine states of $^3H_4$ of $Pr^{3+}$ doped $Y_2SiO_5$ (Pr:YSO, the Pr concentration is 0.05 at. %), and the excited state |3> is selectively chosen from the hyperfine states of $^1D_2$. In Fig. 1(b) noncollinear laser beams are focused onto the sample of Pr:YSO by a 30 cm focal length lens. In the sample of Pr:YSO whose length is 3 mm, all laser beams are nearly uniformly overlapped. The crossing angle among laser beams is ~35 mrad. The Pr:YSO sample is in a helium cryostat keeping temperature at ~6 K. The optical powers of the probe $E_P$, the coupling $\Omega_C$, and the read-out $\Omega_A$ are 0.2 mW, 10 mW, and 15 mW before incidence into the sample, respectively. The probe $E_P$ (the read-out $\Omega_A$) pulse length is 8 μs (10 μs). The data in Fig. 1(c) is 30 sample averaged by a boxcar integrator. The light pulses are made from a single laser source of Coherent Ring-Dye laser system (Model 899) by using acousto-optic modulators driven by microwave synthesizer (PTS 250) and digital delay generators (SRS DDG 535). For the spectral selection of ~MHz, a repump laser is also applied (not shown in Fig. 1): See Refs. 4, 20, and 22.

Figures 1a and 1b respectively show a schematic energy-level diagram and a propagation scheme of the present delayed all-optical routing. Both optical pulses $E_P$ and $\Omega_C$ are used for EIT and slow light. Under the control of $\Omega_C$, the probe pulse $E_P$ can get slowed down with a delay time τ~20 μs as shown in (i) of Fig. 1(c). The pulse length of the $\Omega_C$ is 100 μs centered at the probe $E_P$. During EIT preparation spin coherence between the ground states |1> and |2> becomes gradually excited. When the read-out pulse $\Omega_A$ arrives later at t=20 μs to temporally coincide with the slow light $E_S$ as shown in (v) of Fig. 1c, the slow light $E_S$ disappears while a new light pulse $E_D$ appears as seen in (iii) and (iv): Delayed all-optical routing. This is due to coherence conversion from the spin coherence to the optical coherence via nondegenerate four-wave mixing processes satisfying the phasing matching condition ($\mathbf{k_D}=\mathbf{k_\Omega}-\mathbf{k_P}+\mathbf{k_A}$): Will be discussed in Figs. 2 and 3. The conversion magnitude of the $E_D$ depends on power ratio of $E_P$ and $\Omega_C$. At low power limit of $E_P$, maximum power of $E_D$ occurs at zero detuning, $\Delta=0$. However, at high power limit of $E_P$, it occurs at when $E_P \neq 0$.

From Fig. 1c the delayed routing time of the $E_D$ is determined by the group delay time $\tau_g \sim 20$ μs, and the τ can be simply lengthened by using an optically thicker ($v_g \propto \Omega_C/N$, *where N is number of atoms*) or longer medium ($\tau_g = L/v_g$, *where L is the length of the medium*) to some extent. Surprisingly the delayed all-optical routing process implies an optical buffer memory at an amount of the group delay-time $\tau_g$ of the slow light $E_S$. This time delay of the buffer memory is essential for the slow electronics to perform electro-optical and/or opto-electrical routing/switching processing. The wiggling in (iii) and (iv) of Fig. 1c is not either artifact or noise, but coherence dynamics caused by the $\Omega_A$: Will be discussed in Fig. 3.

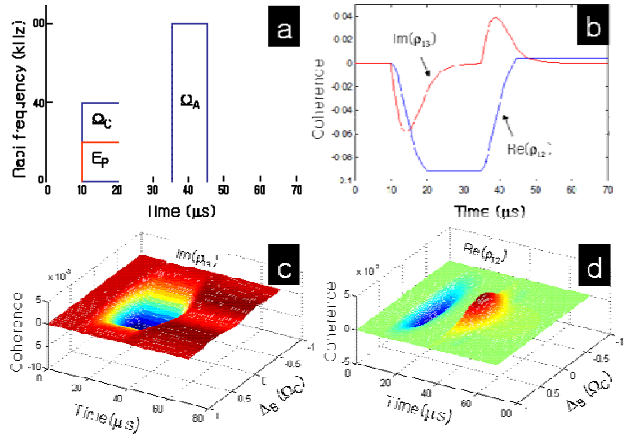

**Figure 2. Numerical simulations of Raman pulse interactions in a spin inhomogeneously broadened model. a**, Pulse sequence; Each pulse length is 10 μs, $\Omega_C$ is twice of $E_P$ in Rabi frequency **b**, Overall optical (Im($\rho_{13}$)) and spin (Re($\rho_{12}$)) coherence. **c**, 2-D simulation of the optical absorption in **b**. **d**, 2-D simulation of the spin coherence in **b**. $\Delta_S$ stands for two-photon detuning between $E_P$ and $\Omega_C$.

For detail discussions of the delayed all-optical routing observed in Fig. 1c, numerical calculations need to be performed. Figure 2a shows a pulse sequence for Fig. 1. Even though the $\Omega_C$ spatio-temporally spans over the slow light $E_S$ in Fig. 1, it has been simplified in Fig. 2a to mimic the actual spin coherence condition propagating with the slow light in the medium. According to the dark state polariton theory [16], the photon component of $E_S$ (Im($\rho_{13}$)) and the spin excitation (Re($\rho_{12}$)) must travel together in both space and time as a combined unity. Because the medium's spin inhomogeneous width is $\Delta_S=30$ kHz in Fig. 1, the excited spin coherence by the $E_P$ and $\Omega_C$ must decay out as of (spin) free induction decay in t=$T_2^{S*}$ ($T_2^{S*}=1/(\pi\Delta_S)$), which is 10 μs: See Ref. 20. However, as mentioned above the excited spin coherence sustains much longer time with the slow light $E_S$ until the spin homogeneous decay time $T_2^S=500$ μs. This is the reason the system of Fig. 1 is simplified as of a homogeneous spin system in Fig. 2a. Figure 2b shows both the spin coherence Re($\rho_{12}$) and photon absorption Im($\rho_{13}$) of $E_P$. At the end of the pulse the excited (both optical and spin) coherence freely evolves. For a very short time scale (t << $T_2^S$) the excited spin coherence Re($\rho_{12}$) changes very little: The spin homogeneous decay time is set $T_2^S=0$ simply for exaggeration. When the read-out pulse $\Omega_A$ is turned on at t=35 μs, the excited spin coherence Re($\rho_{12}$) gradually dies out while photon emission (Im($\rho_{13}$)>0) generates: Coherence conversion (this will be discussed in more detail in Fig. 3).

For better understanding of the coherence conversion process discussed in Fig. 2, we present more detail calculations in Fig. 3. Figure 3 presents coherence calculations for different amplitudes of the read-out pulse $\Omega_A$. Figure 3a shows a pulse sequence with the maximum value of $\Omega_A$. In Fig. 3b the Rabi frequency of the read-out pulse $\Omega_A$ increases at a step of 20 kHz starting from 20 kHz (curves (1) and (i)) to 140 kHz (curves (7) and (vii)). As seen in Fig. 3b the coherence amplitudes of both Re($\rho_{12}$) [curves (1), …, (7)] and Im($\rho_{13}$) [curves (i), …, (vii)] are strongly correlated with each other, and starts to oscillate at above a certain value of the read-out Rabi frequency. This is due to $\Omega_A$ induced Rabi oscillation in a population shelved system (see Ref. 20). Here it should be noted that the photo-diode signal $E_D$ in Fig. 1c should be $I_D=|E_D|^2 \propto [\text{Im}(\rho_{13})]^2$.

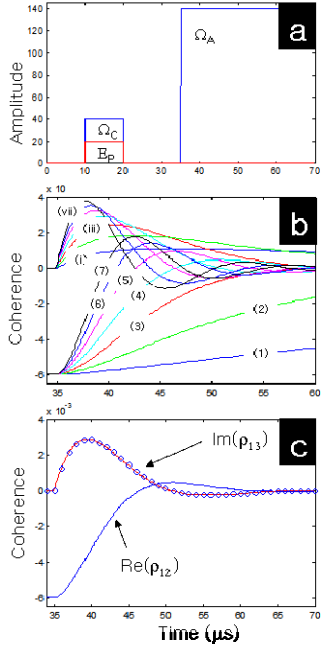

**Figure 3. Coherence conversion via nondegenerate four-wave mixing processes. a**, Pulse sequence. **b**, Optical (i, ii, …, vii) and spin (1, 2, …, 7) coherence evolution under the action of read-out pulse $\Omega_A$ at different Rabi frequencies (20, 40, …, 140 kHz), respectively. **c**, Conversion process of spin coherence (curve 4) into optical coherence (curve iv) in **b**. The open circles are for the best fit curve calculated by Eq. (2) in the text.

Figure 3c demonstrates coherence conversion process between the spin coherence Re($\rho_{12}$) and the optical coherence Im($\rho_{13}$). The time derivative of the EIT-induced spin coherence is

$$\frac{d\rho_{12}}{dt} = -\frac{i}{2\hbar}\Omega_C \rho_{13} + \frac{i}{2\hbar}E_P \rho_{32} - i(\delta_P - \delta_C)\rho_{12} - \gamma_{12}\rho_{12}, \qquad (1)$$

where $\rho_{ij}$ is a density matrix element, and $\delta_P$ ($\delta_C$) is a detuning of $E_P$ ($\Omega_C$) from the resonance frequency for $|1\rangle-|3\rangle$ ($|2\rangle-|3\rangle$): Even two-photon resonance condition between the $E_P$ and $\Omega_C$ is satisfied, i.e., $\omega_P-\omega_C=\omega_{12}$, $\delta_P - \delta_C \neq 0$ because of spin inhomogeneous broadening (see Fig. 1). For Fig. 3c, the following relation intuitively derived for the coherence conversion process by the read-out pulse $\Omega_A$ is applied:

$$E_D(t) \propto -\frac{d}{dt}[\text{Re}(\rho_{12})(t)]. \qquad (2)$$

The result of Eq. (2) is marked as open circles in Fig. 3c. All seven cases for different Rabi frequencies of $\Omega_A$ are tried and checked for perfect matching. This concludes that the nondegenerate four-wave mixing signal $E_D$ is a direct retrieval of the EIT-excited spin coherence Re($\rho_{12}$). Because denser photons of the slow light $E_S$ contributes more in the spin coherence excitation, the coherence conversion process should be enhanced in the slow-light regime [26]. In this case population change between the ground states 1> and 2> is not a necessary condition for the quantum coherence conversion [27].

In conclusion we have demonstrated delayed all-optical routing using slow light phenomenon. The observed time-delayed all-optical routing has potential for quantum optical information processing by allowing an additional function of dynamic optical buffer memory to solve the optical traffic jam caused by the electronics based optical switching devices. To solve the drawback of near unity delay factor (a ratio of delay time to the pulse length), one may envisage increased delay-bandwidth product even in an EIT-based slow light regime [28].

**Acknowledgements**

The author acknowledges that this work was supported by the Creative Research Initiative Program (the Center for Photon Information Processing) by Korean Ministry of Science and Technology, and KOSEF.